# Effects of topological edge states on the thermoelectric properties of Bi nanoribbons

L. Cheng, H. J. Liu[*], J. H. Liang, J. Zhang, J. Wei, P. H. Jiang, D. D. Fan

*Key Laboratory of Artificial Micro- and Nano-Structures of Ministry of Education and School of Physics and Technology, Wuhan University, Wuhan 430072, China*

Using first-principles calculations combined with Boltzmann transport theory, we investigate the effects of topological edge states on the thermoelectric properties of Bi nanoribbons. It is found that there is a competition between the edge and bulk contributions to the Seebeck coefficients. However, the electronic transport of the system is dominated by the edge states because of its much larger electrical conductivity. As a consequence, a room temperature $ZT$ value exceeding 3.0 could be achieved for both *p*- and *n*-type systems when the relaxation time ratio between the edge and the bulk states is tuned to be 1000. Our theoretical study suggests that the utilization of topological edge states might be a promising approach to cross the threshold of the industrial application of thermoelectricity.

Searching for high performance thermoelectric materials is one of the most important topics in the community of materials science. The efficiency of a thermoelectric material is quantified by the dimensionless figure of merit, defined as $ZT = S^2 \sigma T /(\kappa_e + \kappa_l)$, where $S$ is the Seebeck coefficient, $\sigma$ is the electrical conductivity, $T$ is the absolute temperature, and $\kappa_e$ and $\kappa_l$ are the electronic and lattice thermal conductivity, respectively. For conventional thermoelectric materials, the transport coefficients $S$, $\sigma$ and $\kappa_e$ are usually interrelated in a way which makes it very challenging to achieve a higher $ZT$ value. During the past decades, several promising strategies have been proposed to improve the thermoelectric performance which include maximizing the power factor ($S^2\sigma$) through electronic doping and band engineering [1, 2], as well as minimizing the lattice thermal conductivity through phonon scattering [ 3 ]. It is interesting to find that low-dimensional or nano-structured systems combine both advantages and thus are believed to exhibit significantly larger $ZT$ values than their bulk counterparts [4, 5]. Even so, it is still far from the target value ( $ZT > 3.0$) to compete with the traditional energy conversion methods.

---

[*] Author to whom correspondence should be addressed. Electronic mail: phlhj@whu.edu.cn.

Recently, both theoretical predictions [6, 7, 8] and experimental studies [9, 10, 11, 12] have shown that some of good thermoelectric materials such as $Bi_2Te_3$, $Bi_2Se_3$, and $Bi_{1-x}Sb_x$ are also known as topological insulators (TIs). Such new kind of quantum materials have insulating gaps in the bulk but topologically protected gapless surface or edge states on the boundary [13, 14], and they share some similar material features with good thermoelectric materials such as containing heavy elements and having small band gaps. However, whether these two classes of materials are inherently connected remains mysterious and conceptually perplexing. In particular, it is still controversial whether the topologically protected surface/edge states could be utilized to enhance the thermoelectric performance. For example, Ghaemi *et al.* [15] investigated the in-plane transport of $Bi_2Te_3$ and $Bi_2Se_3$ thin films and found that the surface states from top and bottom layers hybridize, and conventional diffusive transport predicts that the tunable hybridization-induced band gap leads to enhanced thermoelectric performance at low temperatures. However, Rittweger *et al.* [16] concluded that a reduction to metallic behavior of the thermopower and electrical conductivity in the $Bi_2Te_3$ films would be expected, which means that the presence of topological surface states may reduce the thermoelectric performance. Osterhage *et al.* [17] showed that the parallel contributing bulk and surface channels tend to cancel each other out, while Sun *et al.* [18] suggested that the thermoelectric performance of $Bi_2Se_3$ can be elevated substantially by utilizations of its gapless conducting surface. It should be noted that most of these works are focused on three-dimensional (3D) TIs with surface states. Recently, Xu *et al.* demonstrated that the topological edge states in an idealized two-dimensional (2D) TI model system lead to large and anomalous Seebeck effects and the $ZT$ value could be improved to be significantly larger than 1 by optimizing the geometric size [19]. It is thus natural to explore the effects of topological edge states on the thermoelectric performance of a realistic 2D TI. In this work, using first-principles calculations combined with Boltzmann transport theory, we explore the possibility to enhance the thermoelectric properties of Bi nanoribbon by utilization the topologically protected edge states. It is expected that our design strategy may prove to be instrumental in the experimental search of high-performance thermoelectric materials and devices.

Our first-principles calculations are performed by using the projector augmented-wave method [20, 21] within the framework of density functional theory (DFT) [22, 23, 24]. The exchange-correlation energy is in the form of Perdew-Burke-Ernzerhof with [25] generalized gradient approximation (GGA). The vacuum space perpendicular to and along the nanoribbon width is set as 20 Å and 25 Å, respectively, so that the nanoribbon and its periodic images can be treated as independent entities. During the structure optimization, the energy cutoff is set as 210 eV and the Brillouin zone is sampled with 1×1×20 Monkhorst-Pack $k$ meshes. Optima atom positions are determined until the magnitude of the force acting on each atom becomes less than 0.01 eV/Å. The spin-orbit coupling is explicitly considered in our calculations to correctly predict the TI nature. The electronic transport coefficients ($S$, $\sigma$ and $\kappa_e$) are derived from the semi-classical Boltzmann theory [26], while the phonon transport coefficient ($\kappa_l$) can be obtained from molecular dynamics simulations [27].

The Bi nanoribbon can be obtained by cutting the Bi (111) monolayer along a particular direction. In the present work, we only consider nanoribbons with armchair edges. Following the notation for graphene nanoribbons [28], the armchair Bi nanoribbon (ABNR) are classified by the number of dimer lines across the ribbon width and are labeled as $N$-ABNR. The initial and optimized structures of 15-ABNR are illustrated in Figure 1(a) and 1(b), respectively. Upon structure relaxations, we see there has an obvious edge reconstruction and all the dangling bonds are eliminated, which is very similar to those found in the BiSb nanoribbons [29].

Figure 2 plots the band structures of ABNR with width of 2.6 nm ($N$=6) and 6.5 nm

($N$=15), where the edge states (marked by blue circles) are identified by the wavefunction projection method [31, 30]. For the 6-ABNR, we see it only possesses trivial edge states and has a direct band gap of 0.24 eV. In contrast, there is a single Dirac cone at the Γ point for the 15-ABNR, which indicates it has topologically protected edge states [6]. This can be also verified by simply counting the times of the edge states that across the Fermi level between time-reversal invariant momenta [6]. In fact, we have calculated the topological invariant number of Bi (111) monolayer by using the so-called "parity method" [6], and confirms that it is a 2D TI [31, 32, 33, 34, 35]. On the other hand, extensive calculations for a series ABNRs find that there is a transition from topological trivial to non-trivial edge states with increasing nanoribbon width. It is thus interesting to obtain the minimum width that required for the observation of non-trivial edge states, which is usually referred to as the penetration depth. Detailed analysis of the band structures reveals that the topological transition occurs at a critical width of ~6.5 nm, which coincides with previously reported by using the model Hamiltonian method [31].

Based on the calculated energy band structures, the electronic transport coefficients of ABNRs can be derived by using the Boltzmann theory and the relaxation time approximation. We first focus on the ABNR with topological trivial edge states ($N$=6). Figure 3 shows the room temperature Seebeck coefficients, electrical conductivity, power factor, and $ZT$ value of 6-ABNR as a function of the carrier concentration. Note here we use a constant relaxation time of $2.38 \times 10^{-14}$ s, which is predicted by using the deformation potential (DP) theory [36, 37] considering the acoustic phonons are the main scattering mechanism. We see from Fig. 3(a) that the Seebeck coefficient $S$ has two peak values at smaller carrier concentration, which is a common feature of semiconducting systems. As for the electrical conductivity $\sigma$, we see from Fig. 3(b) that it becomes almost vanished when the carrier concentration is smaller than $1.0 \times 10^{11}$ /cm$^2$, and increases quickly at the band edges where the absolute value of $S$ is relatively small. Such opposite behaviors indicates that there must be a trade-off between $S$ and $\sigma$ so that the power factor $S^2\sigma$ could be maximized at a particular carrier concentration, as can be found in Fig. 3(c). Indeed, we see from Fig. 3(d) that a maximum $ZT$ value of 0.54 for the *p*-type system and 0.75 for the *n*-type system can be achieved at optimized carrier concentrations.

However, for systems with topologically protected edge states ($N \geq 15$), the generally adopted constant relaxation time is no longer valid. Due to the time-reversal symmetry, backscattering is prevented and carriers moving on the edges would have much longer lifetime. As an alternative, we use the dual scattering time model proposed by Xu *et al*. [19], where the relaxation time of the edge states inside and outside the bulk gap are denoted as $\tau_1$ and $\tau_2$, respectively. Here $\tau_2$ is usually approximated to be the same as that of bulk states ($2.38 \times 10^{-14}$ s), and can be decreased by impurities or disorders. The determination of $\tau_1$ depends on the experimentally measured mean free path and Fermi velocity, which is however not available for our investigated system up to now. For simplicity, $\tau_1$ is set as $1.0 \times 10^{-12}$ s which is in the same order of magnitude of those obtained for 2D TI systems such as HgTe quantum well [38]. In terms of the bulk part ($\sigma_{bulk}$, $\kappa_{e,bulk}$, $S_{bulk}$) and edge part ($\sigma_{edge}$, $\kappa_{e,edge}$, $S_{edge}$), the total transport coefficients of ANBR with topological edge states can be expressed as:

$$\sigma_{total} = \sigma_{bulk} + \sigma_{edge}, \tag{1}$$

$$S_{total} = (S_{bulk}\sigma_{bulk} + S_{edge}\sigma_{edge})/\sigma_{total}, \tag{2}$$

$$\kappa_{e,total} = \kappa_{e,bulk} + \kappa_{e,edge}. \tag{3}$$

Note $\kappa_{e,bulk}$ and $\kappa_{e,edge}$ can be obtained by Wiedemann-Franz law, where the Lorentz number is assumed to be $1.5 \times 10^{-8}$ V$^2$/K$^2$ and $2.4 \times 10^{-8}$ V$^2$/K$^2$ for the bulk and edge states, respectively. This is reasonable since the transport of bulk states is similar to that in a non-degenerate semiconductor, while the topological edge states are metallic in nature. Figure 4 plots the transport coefficients and $ZT$ values of 15-ABNR as a function of carrier concentration, where the contributions from bulk and edge states are also shown for comparison. Since the edge states are protected by time-reversal symmetry, $\tau_1$ is immune to nonmagnetic impurities or disorder, which however tends to reduce $\tau_2$. In principle, one can fine tune the relaxation time ratio $r_\tau = \tau_1/\tau_2$ by controlling the form and/or concentration of nonmagnetic impurities or disorder. Let's take $r_\tau = 100$ as an example. We see from Fig. 4(a) that the absolute value of $S_{bulk}$ can be as high as 1000 μV/K for both *p*- and *n*-type carriers. The value of $S_{edge}$ is relatively smaller but the maximum is still exceeding 300 μV/K. More importantly, we find $S_{edge}$ and $S_{bulk}$ have opposite sign, which is believed to be caused by the unique energy dependence of the edge relaxation time [19]. According to Equation (2), this means there is a competition between the contributions of edge and the bulk states but are dominated by the edge states since $\sigma_{edge}$ is much larger than $\sigma_{bulk}$ (Fig. 4(b)). As a consequence, $S_{total}$ is almost equal to $S_{edge}$. The much larger $\sigma_{total}$ combined with moderate $S_{total}$ leads to enhanced $S^2\sigma$ as shown in Fig 4(c), which would undoubtedly contribute to a higher *ZT* value. Indeed, we see from the Fig. 4(d) that a maximum $ZT_{total}$ of 1.6 and 1.7 can be achieved for *p*- and *n*-type carriers, respectively. These values obviously exceed those of the ABNR with only trivial edge states, and implies that one can significantly enhance the thermoelectric performance of Bi nanoribbons by utilizing their topological edge states in an experimentally feasible way.

It has been found that for the 2D quantum spin Hall (QSH) systems, the relaxation time ratio $r_\tau$ can be tuned to as high as 1000 [38, 39]. To discuss the $r_\tau$ dependence of the *ZT* values, we show in Figure 5 the room temperature *ZT* values of 15-ABNR as a function of carrier concentration at a series of $r_\tau$. We see that as $r_\tau$ increases from 50 to 1000, the optimized *ZT* values increase monotonically. This is reasonable since the edge states give more weighted contribution to the power factor at larger $r_\tau$. If $r_\tau$ of 15-ABNR is fine tuned to be 1000, a maximum *ZT* value of 3.0 and 3.1 can be achieved for *p*- and *n*-type carriers, respectively. Such values are not only significantly larger than those of ABNRs having only trivial edge states, but also represent an important step forward to cross the threshold of the industrial application of thermoelectricity.

We thank financial support from the National Natural Science Foundation (grant No. 11574236 and 51172167) and the "973 Program" of China (Grant No. 2013CB632502).

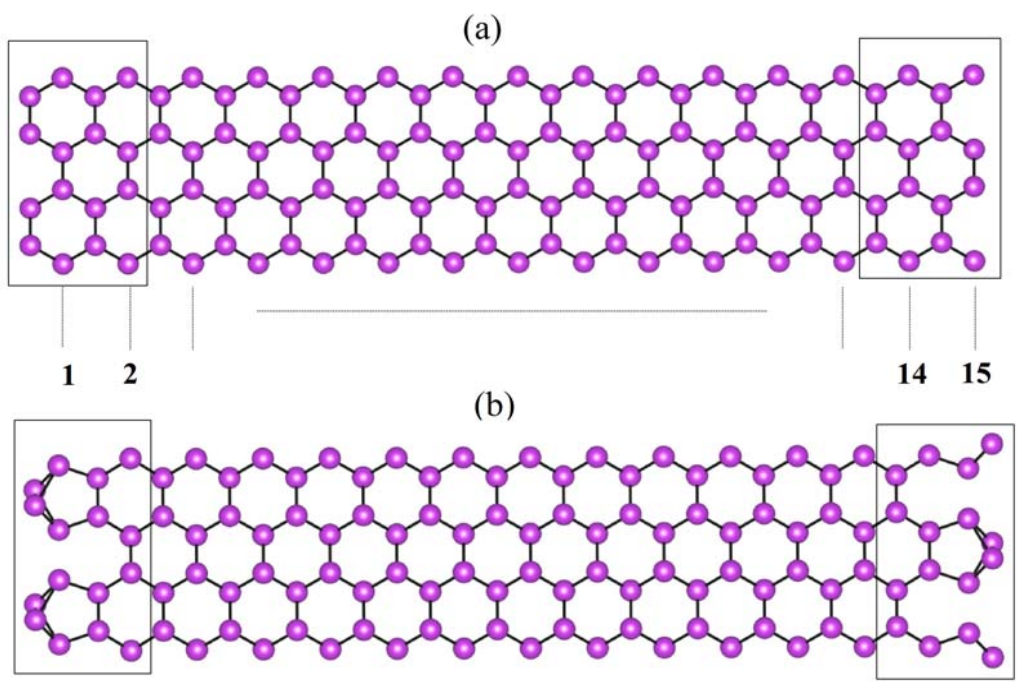

**Figure 1** Ball-and-stick model of (a) the initial, and (b) the fully relaxed structure of 15-ABNR.

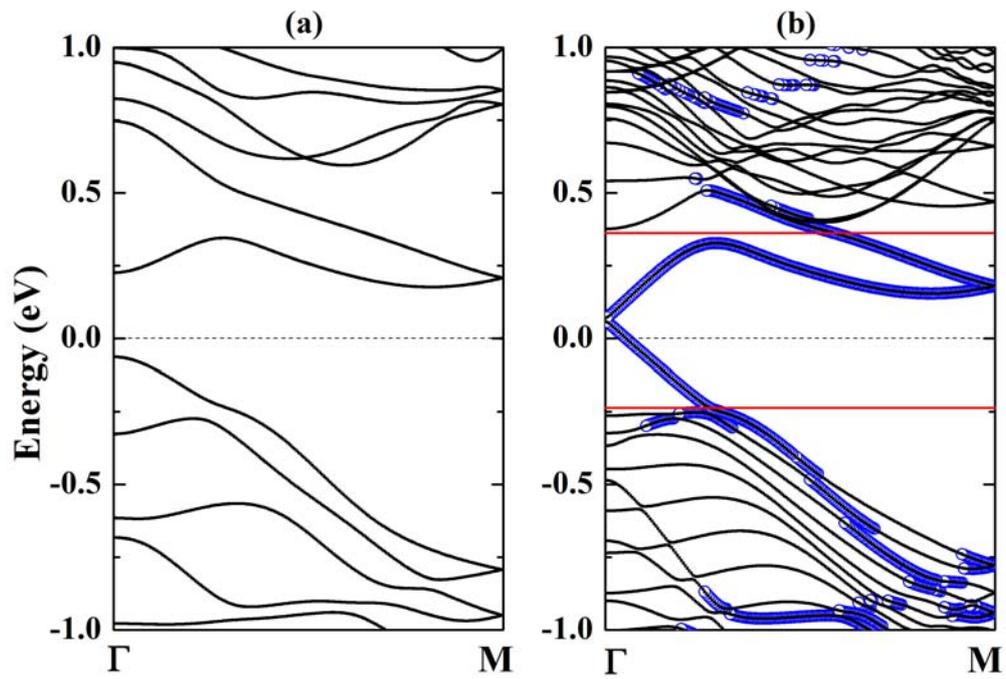

**Figure 2** Band structures of (a) 6-ABNR, and (b) 15-ABNR, where the edge states are marked by blue circles. The Fermi level is at 0 eV, and the red lines indicate the bulk gap.

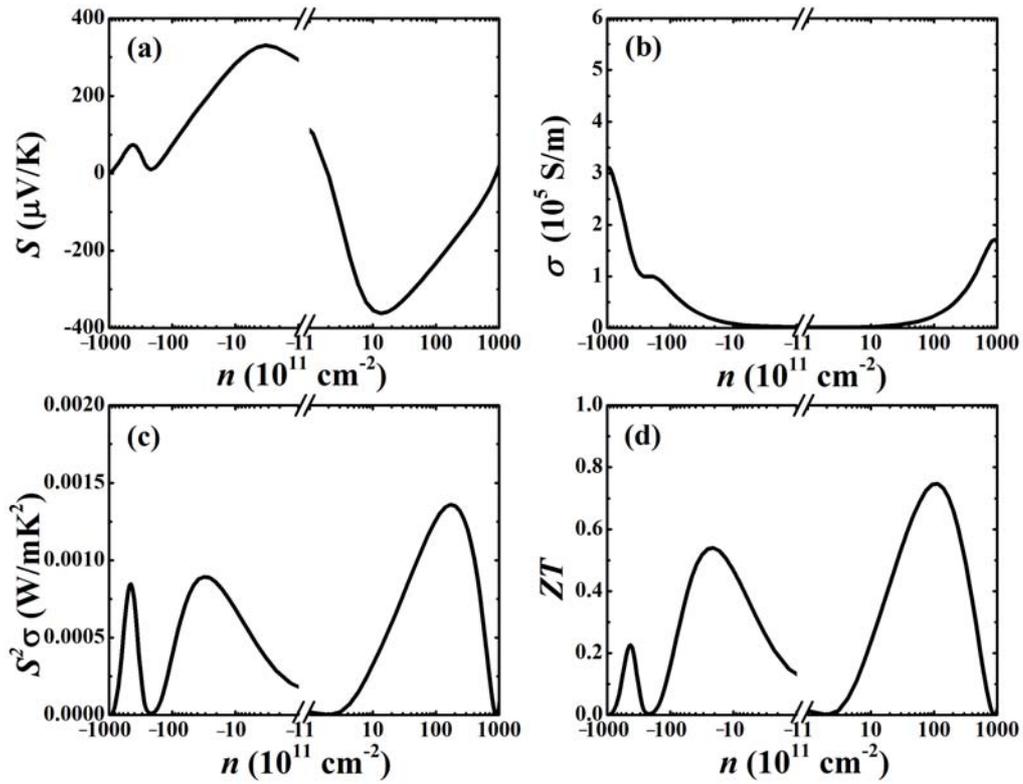

**Figure 3** Room temperature (a) Seebeck coefficient $S$, (b) electrical conductivity $\sigma$, (c) power factor $S^2\sigma$, and (d) $ZT$ values of 6-ABNR, plotted as a function of carrier concentration. Positive and negative carrier concentrations represent $n$- and $p$-type carriers, respectively.

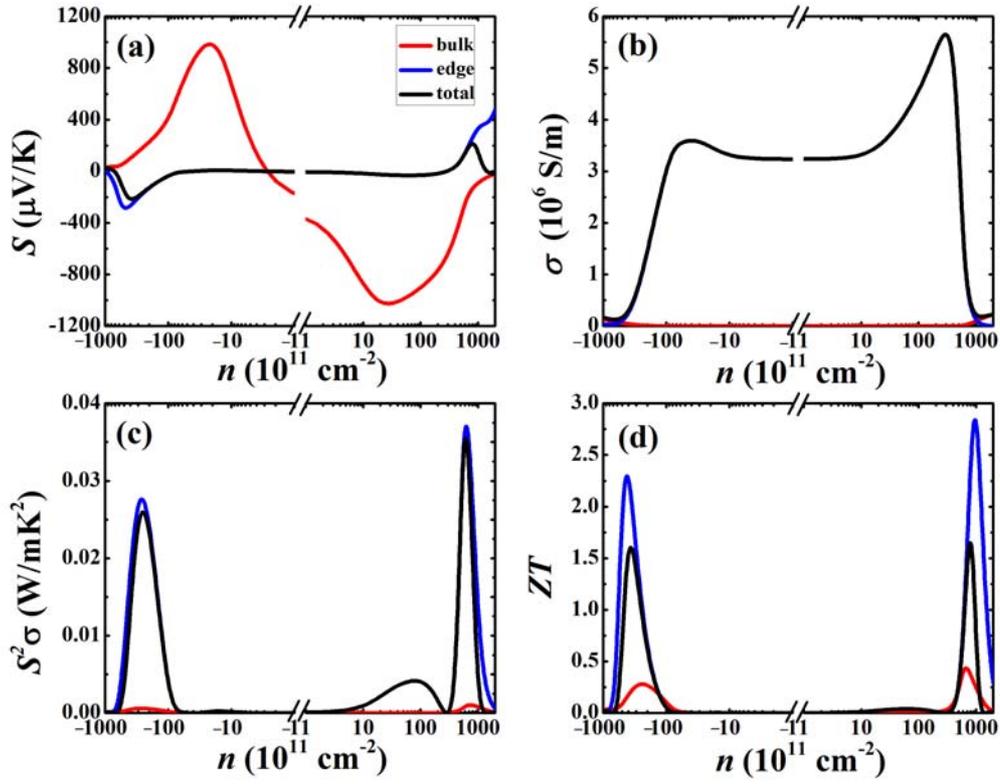

**Figure 4** Room temperature (a) Seebeck coefficient $S$, (b) electrical conductivity $\sigma$, (c) power factor $S^2\sigma$, and (d) $ZT$ values of 15-ABNR, plotted as a function of carrier concentration. Positive and negative carrier concentrations represent $n$- and $p$-type carriers, respectively. The contributions from bulk and edge states are also shown for comparison. The relaxation time ratio $r_\tau$ is set to 100.

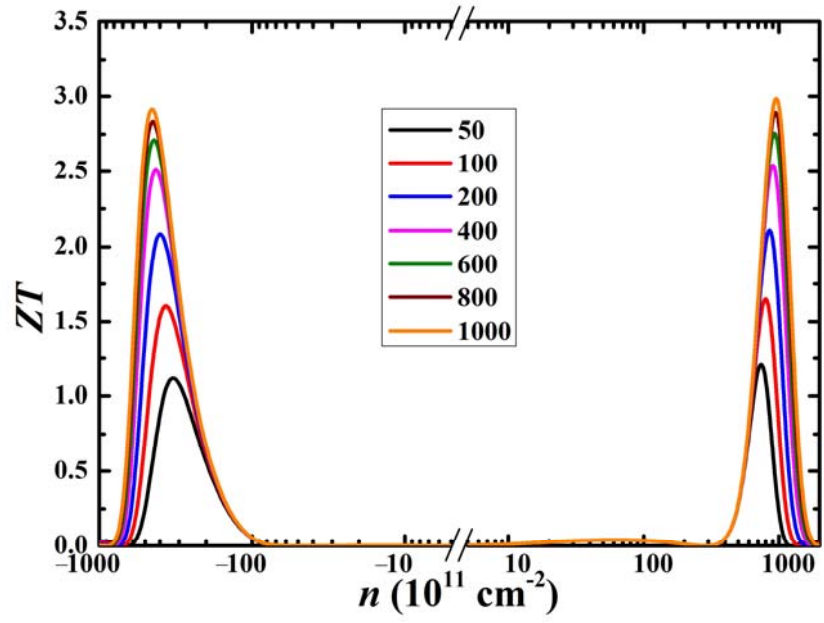

**Figure 5** Room temperature *ZT* values of 15-ABNR as a function of carrier concentrations at a series relaxation time ratio $r_\tau$. Positive and negative carrier concentrations represent *n*- and *p*-type carriers, respectively.

**Reference**

[1] J. P. Heremans, V. Jovovic, E. S. Toberer, A. Saramat, K. Kurosaki, A. Charoenphakdee, S. Yamanaka, and G. J. Snyder, Science **321**, 554 (2008).

[2] Y. Z. Pei, X. Shi, A. LaLonde, H. Wang, L. Chen, and G. J. Snyder, Nature **473**, 66 (2011).

[3] B. Poudel, Q. Hao, Y. Ma, Y. Lan, A. Minnich, B. Yu, X. Yan, D. Wang, A. Muto, D. Vashaee, X. Chen, J. Liu, M. S. Dresselhaus, G. Chen, and Z. Ren, Science **320**, 634 (2008).

[4] L. D. Hicks, and M. S. Dresselhaus, Phys. Rev. B **47**, 12727 (1993).

[5] L. D. Hicks, and M. S. Dresselhaus, Phys. Rev. B **47**, 16631 (1993).

[6] L. Fu, and C. L. Kane, Phys. Rev. B **76**, 045302 (2007).

[7] S. Murakami, New J. Phys. **9**, 356 (2007).

[8] H. Zhang, C. X. Liu, X. L. Qi, X. Dai, Z. Fang, and S. C. Zhang, Nat. Phys. **5**, 438 (2009).

[9] D. Hsieh, D. Qian, L. Wray, Y. Xia, Y. S. Hor, R. J. Cava, and M. Z. Hasan, Nature **452**, 970 (2008).

[10] Y. Xia, D. Qian, D. Hsieh, L. Wray, A. Pal, H. Lin, A. Bansil, D. Grauer, Y. S. Hor, R. J. Cava, and M. Z. Hasan, Nat. Phys. **5**, 398 (2009).

[11] Y. L. Chen, J. G. Analytis, J. H. Chu, Z. K. Liu, S. K. Mo, X. L. Qi, H. J. Zhang, D. H. Lu, X. Dai, Z. Fang, S. C. Zhang, I. R. Fisher, Z. Hussain, and Z. X. Shen, Science **325**, 178 (2009).

[12] D. Hsieh, Y. Xia, D. Qian, L. Wray, F. Meier, J. H. Dil, J. Osterwalder, L. Patthey, A. V. Fedorov, H. Lin, A. Bansil, D. Grauer, Y. S. Hor, R. J. Cava, and M. Z. Hasan, Phys. Rev. Lett. **103**, 146401 (2009).

[13] M. Z. Hasan and C. L. Kane, Rev. Mod. Phys. **82**, 3045 (2010).

[14] X. L. Qi and S. C. Zhang, Rev. Mod. Phys. **83**, 1057 (2011).

[15] Pouyan Ghaemi, Roger S. K. Mong, and J. E. Moore, Phys. Rev. Lett. **105**, 166603 (2010).

[16] F. Rittweger, N. F. Hinsche, P. Zahn, and I. Mertig, Phys. Rev. B **89**, 035439 (2014).

[17] H. Osterhage, J. Gooth, B. Hamdou, P. Gwozdz, R. Zierold, and K. Nielsch, Appl. Phys. Lett. **105**, 123117 (2014).

[18] G. L. Sun, L. L. Li, X. Y. Qin, D. Li, T. H. Zou, H. X. Xin, B. J. Ren, J. Zhang, Y.

Y. Li, and X. J. Li, Appl.Phys. Lett. **106**, 053102 (2015).

[19] Y. Xu, Z. X. Gan, and S. C. Zhang, Phys. Rev. Lett. **112**, 226801 (2014).

[20] P. E. Blöchl, Phys. Rev. B **50**, 17953 (1994).

[21] G. Kresse, and J. Joubert, Phys. Rev. B **59**, 1758 (1999).

[22] G. Kresse, and J. Hafner, Phys. Rev. B **47**, 558 (1993).

[23] G. Kresse, and J. Hafner, Phys. Rev. B **49**, 14251 (1994).

[24] G. Kresse, and J. Furthmuller, Comput. Mater. Sci. **6**, 15 (1996).

[25] J. P. Perdew, K. Burke, and M. Ernzerhof, Phys. Rev. Lett. **77**, 3865 (1996).

[26] G. K. H. Madsen and D. J. Singh, Comput. Phys. Commun. **175**, 67 (2006).

[27] L. Cheng, H. J. Liu, X. J. Tan, J. Zhang, J. Wei, H. Y. Lv, J. Shi and X. F. Tang, J. Phys. Chem. C **118**, 904 (2014).

[28] Y. W. Son, M. L. Cohen, and S. G. Louie, Phys. Rev. Lett. **97**, 216803 (2006).

[29] H. Y. Lv, H. J. Liu, X. J. Tan, L. Pan, Y. W. Wen, J. Shi, and X. F. Tang, Nanoscale, **4**, 511(2012).

[30] K. Park, J. J. Heremans, V. W. Scarola, and D. Minic, Phys. Rev. Lett. **105**, 186801 (2010).

[31] M. Wada, S. Murakami, F. Freimuth, and G. Bihlmayer, Phys. Rev. B **83**, 121310(R) (2011).

[32] Z. Liu, C. X. Liu, Y. S. Wu, W. H. Duan, F. Liu, and J. Wu, Phys. Rev. Lett. **107**, 136805 (2011).

[33] F. Yang, L. Miao, Z. F. Wang, M. Y. Yao, F. F. Zhu, Y. R. Song, M. X. Wang, J. P. Xu, A. V. Fedorov, Z. Sun, G. B. Zhang, C. H. Liu, F. Liu, D. Qian, C. L. Gao, and J. F. Jia, Phys. Rev. Lett. **109**, 016801 (2012).

[34] H. Kotaka, F. Ishii, M. Saito, T. Nagao, and S. Yaginuma, Jpn. J. Appl. Phys. **51** 025201 (2012).

[35] L. Chen, Z. F. Wang, and F. Liu, Phys. Rev. B **87**, 235420 (2013).

[36] J. Bardeen, and W. Shockley, Phys. Rev. **80**, 72 (1950).

[37] J. Y. Xi, M. Q. Long, L. Tang, D. Wang, and Z. G. Shuai, Nanoscale **4**, 4348 (2012).

[38] V. Daumer, I. Golombek, M. Gbordzoe, E. G. Novik, V. Hock, C. R. Becker, H. Buhmann, and L. W. Molenkamp, Appl. Phys. Lett. **83**, 1376 (2003).

[39] G. M. Gusev, Z. D. Kvon, O. A. Shegai, N. N. Mikhailov, S. A. Dvoretsky, and J. C. Portal, Phys. Rev. B **84**, 121302 (2011).